\documentstyle[12pt,fleqn]{article}
\begin{document}
\title{\bf On the relation between metric and spin--2 formulations of
linearized Einstein theory}
\author{Jacek Jezierski\\
Department of Mathematical Methods in
Physics, \\ University of Warsaw,
ul. Ho\.za 74, 00-682 Warsaw,
Poland}
\date{ Ref. No. 4556}
\maketitle

\newcommand{\tendu}[3]{{#1}_{#2}{^{#3}}}
\newcommand{\base}[2]{{{\partial}\over {\partial {#1}^{#2}}}}

\begin{abstract}
A twenty--dimensional space of charged solutions of spin--2 equations is
proposed. The relation with extended (via dilatation) Poincar\'e group  is
analyzed. Locally, each solution of the theory may be described in terms of a
potential, which can be interpreted as a metric tensor satisfying linearized
Einstein equations.
Globally, the non--singular metric tensor exists if and only if 10 among the
above 20 charges do vanish.
The situation is analogous to that in classical electrodynamics, where
vanishing of  magnetic monopole implies the global existence of the
electro--magnetic potentials.
The notion of {\em asymptotic conformal Yano--Killing tensor} is
defined and used as a basic concept to introduce an inertial frame in
General Relativity via asymptotic conditions at spatial infinity.
  The introduced class of
asymptotically flat solutions is free of supertranslation ambiguities.
\end{abstract}

\section{Introduction}
The linearized Einstein equations describing weak gravitational field
can be formulated in
terms of the metric tensor (see Section 2) or, in terms of the Weyl
tensor, as a spin--2 field (see Section 3). Both formulations are globally
equivalent if the topology of the spacetime is trivial.
However, the linearized Einstein theory can only be applied in the
asymptotically flat region, which has a nontrivial topology
(a tube containing the strong field region has to be removed from Minkowski
space). In this case both formulations
are no longer globally equivalent. Similarly as in classical electrodynamics,
where the vanishing of magnetic monopoles in topologically non--trivial regions
implies the existence of magnetic vector--potential (\cite{Dirac},
\cite{Amaldi}), the necessary and
sufficient condition for the equivalence of the two formulations of the
linearized gravity is the vanishing of certain charges which we introduce in
Section 4.

The new charges result in a natural way from a
geometric formulation of the ``Gauss law'' for the gravitational
charges, defined
in terms of the Riemann tensor. We present this formulation in Section 5. It
leads to the notion of the {\em conformal Yano--Killing tensor}.
A conformal Yano--Killing (CYK) equation (\ref{CYK}) posseses
twenty--dimensional space of
solutions for flat Minkowski metric in four--dimensional spacetime ($n=4$).
There is no obvious correspondence between ten--dimensional
asymptotic Poincar\'e
group and the twenty--dimensional space of CYK tensors. Only half of them
(the four--momentum vector $p_\mu$ and the angular momentum
tensor $j_{\mu\nu}$) are Poincar\'e generators.
This situation is analogous to that of electrodynamics,
where, in topologically non--trivial regions, we have two charges
(electric + magnetic) despite the fact that the gauge group is
one--dimensional.

Let us notice, that for $n=2$ ($n$ is a dimension of spacetime) the space
of solutions of the equation (\ref{CYK}) is infinite and for $n=3$ the
corresponding space is only four--dimensional. Possible dimensions we
summarize in a table:\\[2ex]
\begin{tabular}{|r|c|c|c|}
\hline
dimension of spacetime & $n=2$ & $n=3$ & $n=4$ \\
dimension of (pseudo)euclidean group & 3 & 6 & 10 \\
dimension of conformal group & $\infty$ & 10 & 15 \\
dimension of space of CYK tensors & $\infty$ & 4 & 20 \\ \hline
\end{tabular} \\[2ex]
The above table shows that there is no obvious relation
between CYK tensors and the group.

On the other hand, in the case $n=4$, it is possible to connect CYK
tensors with
eleven--dimensional group of Poincar\'e transformations enlarged by dilatation
(pseudo--si\-mi\-la\-ri\-ty transformations). Eleven--dimensional
algebra (space of
Killing vectors) of this group
allows us to construct (via the wedge product) all the CYK tensors
in Minkowski spacetime.

A natural application of the above construction to the description of
asymptotically flat spacetimes is proposed in sections 6 and~7. It
allows us to
define an asymptotic charge at spatial infinity without supertranslation
ambiguities. The existence or nonexistence of the
corresponding asymptotic CYK tensors can be  chosen as a criterion for
classification of asymptotically flat spacetimes.  For example, the
Taub--NUT spacetimes \cite{NUT} can
be excluded assuming that the corresponding
conserved quantity vanishes. Similarly, the Demia\'nski solution \cite{MD}
corresponds to a non--vanishing charge $\bf d$  and can also be excluded this
way (see Section 4).

We stress that the new 10 charges introduced in
the present paper
are different from 10 exact gravitationally--conserved quantities at null
infinity, considered by E.T. Newman and R. Penrose \cite{NP}.
The situation at null infinity (which we do not analyze in our paper) is
much more delicate than the one at spatial infinity. To extend our definition
to the null infinity, we will probably have to assume  additional
asymptotic conditions which guarantee the existence of the asymptotic CYK
tensor.

\section{Linearized gravity}
Linearized
Einstein theory (see e.g. \cite{MTW} or \cite{AT}) can be formulated as
follows.
Einstein equation
 $$2G_{\mu \nu} (g) = 16\pi T_{\mu \nu}$$
give, after linearization:
\begin{equation}  \label{lrE}
h_{\mu \alpha}{^{;\alpha}}{_{\nu}} +
	 h_{\nu \alpha}{^{;\alpha}}{_{\mu}} -
h_{\mu \nu}{^{;\alpha}}{_{\alpha}} -
(\eta ^{\alpha \beta}h_{\alpha \beta}){_{;\mu \nu}} - \eta _{\mu \nu}
[ h_{\beta \alpha}{^{;\alpha \beta}} -
h_{\alpha}{^{\alpha ; \beta}}{_{\beta}} ] = 16\pi T_{\mu \nu} \end{equation}
\noindent
where pseudoriemannian metric $g_{\mu \nu} = \eta _{\mu \nu} + h_{\mu \nu}$,
$\eta _{\mu \nu}$ is the flat Minkowski metric,
 ``$;$'' denotes four--dimensional covariant derivative with respect
to the metric $\eta_{\mu \nu}$.

It is useful to define the following object:
\[ H^{\mu \alpha \nu \beta} := h^{\alpha \nu} \eta^{\mu \beta} +
\eta^{\alpha \nu} h^{\mu \beta} -h^{\mu \nu} \eta^{\alpha \beta}
-\eta^{\mu \nu} h^{\alpha \beta} + h^{\gamma}{_{\gamma}}
( \eta^{\mu \nu} \eta^{\alpha \beta} -\eta^{\alpha \nu} \eta^{\mu \beta})
\] \noindent
which fulfills the following identities:
\[ H^{\mu \alpha \nu \beta}=H^{\nu \beta \mu \alpha} =
H^{[\mu \alpha ] [ \nu \beta ]}
\; \; \; \; H^{\mu [\alpha \nu \beta ]}=0 \]
The equation (\ref{lrE}) may be rewritten as:
\begin{equation}\label{lrE1}
H^{\mu \alpha \nu \beta}{_{;\alpha\beta}} = 16 \pi T^{\mu \nu}
\end{equation}
Let $\Sigma = \{ x^0= \mbox{const.} \}$ be a spacelike hyperplane.
We can define the energy--momentum vector $p^{\mu}$:
\begin{equation}\label{p}
 16\pi p^{\mu} := 16\pi \int_{\Sigma } T^{\mu 0} =
\int_{\Sigma } H^{\mu \alpha 0 \beta}{_{,\alpha\beta}}=
\int_{\Sigma } H^{\mu \alpha 0k}{_{,\alpha k}} =
\int_{\partial \Sigma } H^{\mu \alpha 0k}{_{,\alpha}} {\rm d}^2\! S_k
\end{equation}
and the angular momentum tensor $j^{\mu\nu}$:
\[ 16\pi j^{\mu\nu} := 16\pi \int_{\Sigma } {\cal J}^{\mu\nu 0} =
16\pi \int_{\Sigma } ( x^{\mu}T^{\nu 0}-x^{\nu}T^{\mu 0} )= \]
\begin{equation}\label{J}
=\int_{\partial \Sigma } ( x^{\mu} H^{\nu \alpha 0k}{_{,\alpha}} -x^{\nu}
H^{\mu \alpha 0k}{_{,\alpha}} + H^{\mu k0\nu} - H^{\nu k 0\mu} )
 {\rm d}^2\! S_k \end{equation}
where
\begin{equation}
{\cal J}^{\mu\nu \kappa} :=
x^{\mu}T^{\nu\kappa}-x^{\nu}T^{\mu\kappa}
\; ; \;\;  \ \  {\cal J}^{\mu\nu \kappa}{_{,\kappa}}=0
\end{equation}
Here $(x^\mu )$ is a global (pseudo-)cartesian coordinate system
in the Minkowski space. Without coordinates, we may relate the above
quantities to the Poincar\'e group generators:
\[ {\cal T}_{\mu}=\base x\mu \; , \; \; \; {\cal L}_{\mu\nu}=x_{\mu} \base
x\nu - x_\nu \base x\mu \; \; \; \; (\mbox{where} \;
x_\mu=\eta_{\mu\nu}x^\nu ) \] \noindent
The index ``$\mu$'' in (\ref{p}) refers to a
translation Killing field $\cal T$ and ``$\mu\nu$'' in (\ref{J}) refers
respectively  to a generator $\cal L$ of proper
Lorentz transformations.

We have also conservation laws:
\[ \frac d{dt} p^{\alpha} +\int_{\partial \Sigma } T^{\alpha k} {\rm
d}^2\! S_k =0 \]
\[ \frac d{dt} j^{\mu\nu} +\int_{\partial \Sigma } {\cal J}^{\mu\nu k}
{\rm d}^2\! S_k =0 \]

Equation (\ref{lrE}) is invariant with respect to the ``gauge''
transformation:
\begin{equation}
	h_{\mu \nu} \rightarrow h_{\mu \nu} + \xi _{\mu ; \nu} +
\xi _{\nu ; \mu}   \label{gauge4}
\end{equation}
where $\xi _{\mu}$ is a covector field.

The 3+1 decomposition of (\ref{lrE}) gives 6
dynamical equations for the space--space components $h_{kl}$ of the
metric (latin indices run from 1 to 3) and 4 equations which do not
contain time derivatives of $h_{kl}$. It is convenient to introduce the
``ADM--momentum'' $P^{kl}$, by the following formula (for full nonlinear
theory see \cite{ADM}):
\begin{equation}
	2\Lambda ^{-1}P_{kl}  = \dot{h} _{kl} -(h_{0k | l} + h_{0l | k})
+ \eta_{kl}(2h_0{^m}{_{|m}} -\dot h )  \label{defP}
\end{equation}
where $h := \eta_{kl}h^{kl}$ ,
$\Lambda :=(\det \eta_{kl})^{1/2} $, the symbol ``$|$'' denotes the
three--dimensional covariant derivative and dot means, as usual, the
time derivative. This way the 6 dynamical, second order equations
can be written as the system of 12 first order (in time) equations:
\begin{equation}
 \dot{P} _{kl} = \frac{\Lambda}2 \left(
 h_{kl}{^{| m}}{_m} + h_{| kl}  -
h^m{_{k | ml}} - h^m{_{l | km}} + h^0{_{0 | kl}} - \eta_{kl}
h^0{_0}{^{| m}}{_m} \right) + 8\pi \Lambda T_{kl} \label{Pdot}
\end{equation}
 \begin{equation}
	\dot{h} _{kl} = 2\Lambda ^{-1}(P_{kl} - \frac{1}{2}\eta_{kl}P)
   + h_{0k | l} + h_{0l | k}	\label{hdot}
\end{equation}
where $P = \eta^{kl}P_{kl}$.\\
The constraint equations can be written as follows:
\begin{equation}
  P^{kl}{_{| l}} = -8\pi \Lambda T^{0k} 
	\label{conP}
\end{equation}
\begin{equation}
   h^{kl}{_{| kl}} - h^{| k}{_k} = 16\pi T^{00}
   	\label{conh}
\end{equation}
Equations (\ref{conP}) and (\ref{conh}) are called the vector constraint and
the scalar constraint respectively.
As a consequence of (\ref{gauge4}) and (\ref{defP}) the gauge splits
also into its time--like component $\xi _0$ which acts on $P_{kl}$:
\[	\Lambda ^{-1} P_{kl} \rightarrow \Lambda ^{-1} P_{kl} - \xi _{0 |
kl} + \eta_{kl} \xi _0{^{| m}}{_m}
 \] \noindent
and a three--dimensional gauge $\xi_k$ acting on the three--metric:
\[
	h_{kl} \rightarrow h_{kl} + \xi_{l | k}+ \xi_{k | l}
\]
The Cauchy data ($g_{kl}$, $P^{kl}$) in $\Sigma $ are equivalent to
($\overline{g}_{kl}$, $\overline{P}^{kl}$) if they can be related by the
gauge transformation with $\xi _{\mu}$ vanishing on $\partial \Sigma $.
 The evolution of canonical
variables $P^{kl}$ and $h_{kl}$ given by equations (\ref{Pdot} -- \ref{hdot})
 is not unique unless the lapse function ($h^0{_0}$) and the shift vector
($h^0{_k}$) are specified.

To describe the dynamics we take as an example the volume $V \subset
\Sigma$ contained
between the two spheres of radius $r_0$ and $r_1$ respectively:
\[
V=K(0,r_0,r_1)
\] \noindent
We are interested in exterior vacuum solutions so we are assuming that
$T_{\mu\nu}=0$ in $V$.
Limiting cases $r_0 \rightarrow 0$ and/or
 $r_1 \rightarrow \infty$ will also be considered.
We use radial coordinates in $V$: $y^3=r$,
$y^1=\theta $, $y^2=\phi $ (spherical angles ($y^A$) $A=1,2$).
Moreover, $y^0=x^0$ denotes the time coordinate.

The entire gauge--independent information about the state
 of the gravitational field ($h_{kl},\ P^{kl}$) is contained
  in the Riemann tensor $R_{\mu \nu \lambda \kappa }$
   which, in linear approximation, can be expressed
in the following way by $h_{\mu\nu}$

\begin{equation}\label{rieman}
2R_{\mu \nu \lambda \kappa }= h_{\mu \kappa ;\nu \lambda} + h_{\nu \lambda
;\mu \kappa} - h_{\nu \kappa ;\mu \lambda} -h_{\mu \lambda ;\nu \kappa}
\end{equation}

Let ${\bf a}$ denotes the two--dimensional Laplace--Beltrami operator on a
unit sphere $S(1)$. Moreover, $H:=\eta^{AB}h_{AB}$, $\chi_{AB}:= h_{AB}
-\frac 12 \eta_{AB} H$, $S:=\eta^{AB}P_{AB}$, $S_{AB}:= P_{AB}
-\frac 12 \eta_{AB} S$ according to the notation used in \cite{GRG}.

 The following four objects built from the
Riemann tensor can be expressed in terms of the Cauchy data:
\begin{equation}
\overline{\bf x} = r^2{\eta}^{AC}{\eta}^{BD}R_{ABCD} =
 2h^{33}+2rh^{3C}{_{||C}}+r^2\chi^{AB}{_{||AB}}
	-(rH),{_3}-\frac{1}{2}{\bf a}H
		\label{x}
\end{equation}
\begin{equation}
\overline{\bf X}= 2\Lambda r^2{\eta}^{AC}{\eta}^{BD}R_{0DAB;C} =
 2r^2S^{AB}{_{||AB}} + 2rP^{3A}{_{||A}} +{\bf a}P^{33}
	 	\label{X}
\end{equation}
\begin{equation}
\overline{\bf y}=r^2{\varepsilon}^{AC} R_{03AC}  =
2\Lambda^{-1}r^2P^{3A||B}\varepsilon_{AB} \label{y}
\end{equation}
\begin{equation}
\overline{\bf Y}=2\Lambda r^2{\varepsilon}^{AC}{\eta}^{BD}R_{3BCD;A}=
\Lambda({\bf a}+2)h^{3A||B}\varepsilon_{AB}
	 - r^2(\Lambda \chi^C{_{A||CB}}\varepsilon^{AB}),{_3} \label{Y}
\end{equation}
where on each sphere $S(r)$ we denote
by $\varepsilon ^{AB}$ the Levi--Civita
antisymmetric tensor such that $\Lambda
\varepsilon ^{12} =1$ and by ``$||$'' the two--dimensional covariant
derivative related to the two--metric $\eta_{AB}$.

We are ready now to rewrite equations (\ref{p}) and (\ref{J}):

\begin{equation}
16\pi p^0=\int_{\partial V} \Lambda (h^{3k}{_{|k}}-h^{|3})=
\int_{\partial V} \frac{\Lambda}r \left[ 2h^{33} -(rH),{_3} \right] =
\hspace{0.5cm} \int_{\partial V} \frac{\Lambda}r \overline{\bf x}
\end{equation}
\[
16\pi p^z = -2\int_{\partial V} P^{3z} =
 -2\int_{\partial V} \left[ P^{33}\cos\theta +P^{3A}(r\cos\theta )_{,A}
\right]=
\]
\begin{equation}= 2\int_{\partial V} \left( rP^{3A}{_{||A}}
-P^{33}\right)\cos\theta
 =\hspace{0.5cm} \int_{\partial V}  \overline{\bf X} \cos\theta
  \end{equation}
\[
16\pi s^z:= 16\pi j^{xy} = -2\int_{\partial V} P^3{_\phi } =
 -2\int_{\partial V} P^3{_A} (r^2\varepsilon^{AB}\cos\theta)_{||B} =\]
\begin{equation}
 = 2\int_{\partial V} r^2 P^{3}{_{A||B}}\varepsilon^{AB} \cos\theta
 =\hspace{0.5cm} \int_{\partial V} \Lambda \overline{\bf y} \cos\theta
 \end{equation}
\begin{equation}
16\pi j^{z0} +16\pi t p^z = \int_{\partial V} \Lambda \left(
r H^{0k03}{_{|k}} \cos\theta - H^{030z} \right) = \hspace{0.5cm}
\int_{\partial V} \Lambda \overline{\bf x} \cos\theta
\end{equation}
where $(x,y,z)$ are cartesian coordinates on $\Sigma$ ($x=r\sin \theta
\cos\varphi$, $y =r\sin\theta\sin\varphi$, $z=r\cos\theta$) and we show
only typical components of four--vector $p^\mu$ and tensor $j^{\mu\nu}$.

We decompose the four objects $\overline{\bf x},\overline{\bf
y},\overline{\bf X},\overline{\bf Y}$  into the monopole part, the dipole
part and the radiation (wave) part. It can be easily checked that constraint
equations (\ref{conP}) and (\ref{conh}) imply a specific radial dependence
of the monopole and the dipole part:

\begin{equation}  \label{xdot}
	r^2{\eta}^{AC}{\eta}^{BD}R_{ABCD} =
 \frac{4{\bf m}}{r} +
 \frac{12{\bf k}}{r^2} +{\bf x}
\end{equation}

\begin{equation} \label{Xdot}
	2\Lambda r^2{\eta}^{AC}{\eta}^{BD}R_{0DAB;C} =
 \Lambda \frac{12{\bf p}}{r^2}  + {\bf X}
\end{equation}

\begin{equation}\label{ydot}
r^2{\varepsilon}^{AC} R_{03AC}  =
	 \frac{12{\bf s}}{r^2}  +{\bf y}
\end{equation}

\begin{equation}\label{Ydot}
2\Lambda r^2{\varepsilon}^{AC}{\eta}^{BD}R_{3BCD;A}={\bf Y}
 \end{equation}
where ${\bf x}$, ${\bf X}$, ${\bf y}$, ${\bf Y}$ are monopole-- and
dipole--free,
${\bf m}$ is a number (which we identify with the monopole function on
S(1)) and ${\bf k}$, ${\bf p}$, ${\bf s}$
are three--dimen\-sio\-nal vectors
which we identify with the dipole functions on S(1).
Let us notice that the monopole part of $\overline{\bf y}$ and dipole
part of $\overline{\bf Y}$ vanish because of the last equalities in
(\ref{y}) and (\ref{Y}) but not from the definition in terms of the
Riemann tensor. This observation will be analyzed in section 3.

Field equations given in Appendix
imply the following time dependence of the charges:
\[ \dot{\bf m} = \dot{\bf p}=\dot{\bf s}=0 \]
\[ \dot{\bf k} = {\bf p} \]

The dynamical part of the field is described
by the four radiative degrees of freedom ${\bf x},{\bf X},{\bf y},{\bf Y}$.
We proved in \cite{GRG} that Einstein equations are equivalent
to the following dynamical equations:
\[
	\dot{\bf X} = \Lambda \triangle {\bf x}
\]
\[
	 	\dot{\bf Y} = \Lambda \triangle {\bf y}
\]
\[
	\Lambda \dot{\bf x} ={\bf X}
\]
\[
	\Lambda \dot{\bf y} ={\bf Y}
\] \noindent
(where by $\triangle$ we denote the three--dimensional Laplacian) or --
equivalently --  to the pair of wave equations for the ``true degrees of
freedom'' ${\bf x}$  and ${\bf y}$.
We have shown in \cite{GRG} that the variables $\bf x, X, y, Y$ contain
the entire gauge--free information about the unconstrained degrees
of freedom.

\section{Spin--2 field}
We summarize the standard formulation of a spin--2 field $W_{\mu \alpha
\nu \beta}$. This field can be interpreted as a Weyl tensor of linearized
gravity (see \cite{Ch-Kl} or \cite{TF}).
\[ W_{\mu \alpha \nu \beta}=W_{\nu \beta \mu \alpha} =
W_{[\mu \alpha ] [ \nu \beta ]}
\; , \; \;  W_{\mu [\alpha \nu \beta ]}=0 \; , \; \;
W_{\alpha \beta}=W^{\mu}{_{\alpha \mu \beta}} =0 \] \noindent
The above properties define spin--2 field.

The $*$--operation has the following properties:
\[ ({^*} W)_{\alpha\beta\gamma\delta}=\frac 12 \varepsilon_{\alpha \beta
\mu \nu} W^{\mu\nu}{_{\gamma\delta}} \]
\[ (W^*)_{\alpha\beta\gamma\delta}=\frac 12 W_{\alpha\beta}{^{\mu\nu}}
 \varepsilon_{\mu \nu \gamma\delta}  \]
\[ ({^*} W^*)_{\alpha\beta\gamma\delta}=\frac 14 \varepsilon_{\alpha \beta
\mu \nu} W^{\mu\nu\rho\sigma}\varepsilon_{\rho\sigma\gamma\delta} \]

 \[ {^*} W =W^* \; , \; \; {^*} ({^*} W) = {^*}W^* =-W \]
\noindent and ${^*}W$ is called dual spin--2 field.

Field equations:
\[ \nabla_{[\lambda} W_{\mu\nu ] \alpha\beta} =0  \] \noindent
are equivalent to
\[ \nabla^\mu W_{\mu\nu\alpha\beta} =0 \] \noindent
or
\[ \nabla_{[\lambda} {^*}W_{\mu\nu ] \alpha\beta} =0 \] \noindent
or
\[ \nabla^\mu \, {^*}W_{\mu\nu\alpha\beta} =0   \] \noindent
We have already assumed that $T^{\mu\nu}=0$, so the Riemann tensor
$R_{\mu\nu\alpha\beta}$ is equal to the Weyl tensor $W_{\mu\nu\alpha\beta}$.
Let us introduce the following special solution of the above equations
for which the radiative degrees of freedom $\bf x$, $\bf X$, $\bf y$, $\bf Y$
vanish:
\[
\frac{4{\bf m}}{r} +\frac{12{\bf k}}{r^2}=\overline{\bf x} \quad
\Lambda \frac{12{\bf p}}{r^2}  = \overline{\bf X} \quad
\frac{12{\bf s}}{r^2} = \overline{\bf y} \quad
 0 =\overline{\bf Y}
 \]
\[ W_{BC0A}= -\frac 3{r^2} \varepsilon_{BC} \left( \frac{{\bf s},_A}{r}
-\tendu{\varepsilon}AD {\bf p},_D \right)
\]
\[ W_{AB03}= \frac 6{r^4} \varepsilon_{AB} {\bf s} \]
\[ W_{3A30}= \frac 3{r^2} \left( \frac{\tendu{\varepsilon}AD {\bf s},_D}{r}+
{\bf p},_A \right) \]
\[ W_{3AB0}= \frac 3{r^4} \varepsilon_{AB} {\bf s} \]
\[ W_{3030}= -\frac 2{r^3} \left({\bf m}+\frac{3{\bf k}}{r} \right)  \]
\[ W_{0A03}= \frac 3{r^3} {\bf k},_{A}  \]
\[ W_{ABCD}= \frac 2{r^3} \left({\bf m}+\frac{3{\bf k}}{r} \right) \left(
\eta_{AC} \eta_{BD}-\eta_{AD}\eta_{BC} \right) \]
\[ W_{3AB3}=-W_{0AB0} = \frac{\eta_{AB}}{r^3} \left({\bf m}+\frac{3{\bf k}}{r}
\right) \]
\[ W_{BC3A}= -\frac 3{r^3} \varepsilon_{BC} \tendu{\varepsilon}{A}{D} {\bf
k},_{D}\]
where $W_{\mu \nu \lambda \kappa}$ is a spin--2 tensor in the flat
Minkowski space.

It can be easy verified that the following metric tensor $h_{\mu\nu}$
(``potential'') gives the above spin--2 field (as a solution of equation
(\ref{rieman}))

\[ h_{00}={2{\bf m}\over r} + {2{\bf k}\over r^2} \]
\[ h_{0A}= -6{\bf p},_{A} - \frac 2r \tendu{\varepsilon}{A}{B} {\bf s},_{B} \]
\[ h_{03}=-{6{\bf p}\over r}  \]
\begin{equation}\label{metric}
 h_{33}={2{\bf m}\over r} + {6{\bf k}\over r^2} \end{equation}
or in cartesian coordinates $(x^k)$:
\[ h_{00}={2{\bf m}\over r} + {2{\bf k}_m x^m\over r^2} \]
\[ h_{0k}= -{6{\bf p}_k\over r} - \frac 2{r^3} \varepsilon_{klm} {\bf
s}^l x^m \]
\begin{equation}\label{metric1}
 h^{kl}={x^k x^l\over r^2}\left({2{\bf m}\over r} + {6{\bf k}_m x^m\over
r^3}\right) \end{equation}
Applying linearized Einstein equation (\ref{lrE})
to the metric tensor (\ref{metric1}) we obtain the
corresponding  energy--momentum tensor
as a distribution located in origin:
\[ T^{00}={\bf m} \delta - {\bf k}^m \delta ,_m  \]
\[ T^{0k}={\bf p}^k\delta + \frac 12 \epsilon^{kml} {\bf
s}_l \delta ,_m  \]
\begin{equation}\label{tep}
 T^{kl}=0 \end{equation}
where by $\delta$ we have denoted the three--dimensional Dirac's delta and
$\epsilon^{kml}$ is a three--dimensional antisymmetric tensor
($\epsilon^{xyz}=1$).

\section{New charged solution}
The theory presented in the previous Section has interesting charged solutions
which do not admit any global metric tensor as a ``potential''. This situation
is
very similar to the one in  electrodynamics where the
 magnetic monopole solution does not admit any  nonsingular, global vector
potential.

Without assuming the existence of the ``global potential'' $h_{\mu\nu}$ for
the field $W_{\mu\nu \alpha\beta}$,
 we can have a non--vanishing monopole charge ${\bf b}$
in $\overline{\bf y}$
and a non--vanishing dipole charge ${\bf d}$ in $\overline{\bf Y}$.
We have the following nonvanishing
components of the Weyl tensor:
\[ \frac{4{\bf b}}{r} =\overline{\bf y} \quad
\frac{12{\bf d}}{r^2} =\overline{\bf Y} \]

\[ W_{CD3A}= \frac 3{r^2} \varepsilon_{CD} {\bf d},_A \]

\[ W_{0A30}=-\frac 3{r^2} \tendu{\varepsilon}AC {\bf d},_C \]

\[ \dot{\bf s}={\bf d} \]

\[ W_{B03A}= \frac{{\bf b}}{r^3} \varepsilon_{AB} \]

\[ W_{AB03}= \frac{2{\bf b}}{r^3} \varepsilon_{AB}  \]

\[ \dot{\bf b}=0  \] \noindent
The above monopole charge $\bf b$ corresponds to the so called
``NUT charge'' (see \cite{AR}). It is also called the
dual mass (see \cite{CZ}). Both charges satisfy the Gauss law:
\[ \int_{S(r)} (r\overline{\bf y}),_3 = 0 \] \noindent
as a consequence of
\[ (r\overline{\bf y}),_3 + r^3 \varepsilon^{CD} W_{CD0A||B} \eta^{AB} =0
\] \noindent
and
\[ \int_{S(r)} (\overline{\bf Y} \cos\theta),_3 = 0 \] \noindent
from
\[ (\overline{\bf Y}),_3 + 2r^2 \Lambda
 \varepsilon^{BC} W^{3A}{_{B3||CA}} =0 \] \noindent
The potentials can, however, be introduced locally. If we want to
extend their definition for the entire spacetime, we end up
necessarily with a ``wire singularity''. The nonvanishing component of
such a singular
metric for the above monopole solution equals:
\[ h_{0\phi}=4{\bf b}\cos \theta  \] \noindent
If we choose the dipole $\bf d$ parallel to the $z$--axis i.e. ${\bf
d}=d\cos \theta$ then the nonvanishing component of the corresponding
singular metric
for the dipole solution is the following:
\[ h_{\theta\phi}=2rd \sin \theta \cos\theta  \quad \mbox{or} \quad
h_{r\theta}=2d ( \sin^2 \theta \ln\tan\frac{\theta}{2}
 -\cos\theta )
\] \noindent
The above form of the potential $h_{\mu\nu}$ corresponds to the linearized
part of the Demia\'nski solution \cite{MD} or rather to the ``special
Demia\'nski solution'' and $d$ is the so called Demia\'nski parameter
$c$ (see \cite{AK} on pp. 172--173). We propose to call the dipole $\bf
d$ the Demia\'nski charge. \\
We can also introduce two more dipole charges $\bf q$ and $\bf w$ in
$\overline{\bf y}$ and $\overline{\bf x}$  respectively:
\[ 6{\bf q} = \overline{\bf y} \quad
   6{\bf w} =\overline{\bf x} \]
\[ W_{BC0A}= \frac 3{2r} \varepsilon_{BC} {\bf q},_A \]
\[ W_{AB03}= \frac 3{r^2} \varepsilon_{AB} {\bf q} \]
\[ W_{3A03}= \frac 3{2r} \tendu{\varepsilon}AD {\bf q},_D \]
\[ W_{3AB0}= \frac 3{2r^2} \varepsilon_{AB} {\bf q} \]
\[ \dot{\bf d}= -{\bf q} = \ddot{\bf s} \]
\[ \dot{\bf q}=0 \]
\[ W_{3030}=  -\frac{3{\bf w}}{r^2}  \]
\[ W_{0A03}=  -\frac{3}{2r}{\bf w},_A \]
\[ W_{ABCD}=  \frac{3{\bf w}}{r^2}
\left(\eta_{AC} \eta_{BD}-\eta_{AD}\eta_{BC} \right) \]
\[ W_{3AB3}=-W_{0AB0} =  \frac{3{\bf w}}{2r^2}\eta_{AB}  \]
\[ W_{BC3A}= \frac{3}{2r}\varepsilon_{BC}\varepsilon_{A}{^D}{\bf w},_D \]
\[ \dot{\bf p}=-{\bf w} =\ddot{\bf k} \]
\[ \dot{\bf w}=0 \] \noindent
The charges ${\bf q}$ and ${\bf w}$ correspond to the metric tensors which
do not  vanish at spatial infinity ($h_{\mu\nu}=O(1)$).

The fully ``charged'' solution has the following form:
\[
6{\bf w} +\frac{4{\bf m}}{r} +\frac{12{\bf k}}{r^2}=\overline{\bf x} \quad
\Lambda \frac{12{\bf p}}{r^2}  = \overline{\bf X} \]
\[
6{\bf q}+\frac{4{\bf b}}{r} + \frac{12{\bf s}}{r^2} = \overline{\bf y} \quad
\Lambda \frac{12{\bf d}}{r^2} =\overline{\bf Y}
 \]
\[ W_{BC0A}=\varepsilon_{BC}\left( \frac 3{2r}  {\bf q},_A
 + \frac 3{r^2}\tendu{\varepsilon}AD {\bf p},_D
  - \frac 3{r^3}{\bf s},_A \right)
\]
\[ W_{AB03}= \varepsilon_{AB}\left( \frac{3{\bf q}}{r^2}
+ \frac{2{\bf b}}{r^3} +\frac{6{\bf s}}{r^4}   \right) \]

\[ W_{3A30}= -\frac 3{2r} \tendu{\varepsilon}AD {\bf q},_D
+\frac 3{r^2}{\bf p},_A + \frac 3{r^3} \tendu{\varepsilon}AD {\bf s},_D
 \]

\[ W_{3AB0}= \varepsilon_{AB} \left( \frac{3{\bf q}}{2r^2}
 +\frac{{\bf b}}{r^3} + \frac{3{\bf s}}{r^4}   \right) \]

\[ W_{3003}= \frac{3{\bf w}}{r^2}
 +\frac{2{\bf m}}{r^3} +\frac{6{\bf k}}{r^4}  \]

\[ W_{A003}= \frac{3}{2r}{\bf w},_A -\frac 3{r^3} {\bf k},_{A}
-\frac 3{r^2} \tendu{\varepsilon}AC {\bf d},_C \]

\[ W_{ABCD}=\left( \frac{3{\bf w}}{r^2} + \frac{2{\bf m}}{r^3} +
\frac{6{\bf k}}{r^4} \right)
 \left( \eta_{AC} \eta_{BD}-\eta_{AD}\eta_{BC} \right) =
 \left( \frac{3{\bf w}}{r^2} + \frac{2{\bf m}}{r^3} +
\frac{6{\bf k}}{r^4} \right) \varepsilon_{AB}\varepsilon_{CD} \]

\[ W_{3AB3}=-W_{0AB0} =\eta_{AB}\left( \frac{3{\bf w}}{2r^2} +
\frac{{\bf m}}{r^3}+\frac{3{\bf k}}{r^4} \right) \]

\[ W_{3ABC}=\varepsilon_{BC}\left(
 \frac{3}{2r}\varepsilon_{A}{^D}{\bf w},_D + \frac 3{r^2} {\bf d},_A
- \frac 3{r^3} \tendu{\varepsilon}{A}{D} {\bf k},_{D} \right)  \]

In terms of the so--called ``electromagnetic'' tensors (\cite{Ch-Kl}):
\[ E_{kl} = W_{0k0l} \quad H_{kl} = \frac 12 W_{0kij}\epsilon^{ij}{_l} \quad
W_{klmn} = -\epsilon_{kl}{^i}\epsilon_{mn}{^j}E_{ij} \] \noindent
our solutions take the following form:
\[ E_{33}=W_{0303}= - \left( \frac{3{\bf w}}{r^2} + \frac{2{\bf m}}{r^3} +
\frac{6{\bf k}}{r^4} \right) \]
\[ E_{3A}=W_{0A03}= -  \frac{3}{2r}{\bf w},_A + \frac{3}{r^2}
\varepsilon_{A}{^C}{\bf d},_C + \frac{3}{r^3}{\bf k},_A  \]
\[ E_{AB}=W_{0A0B}= \frac 12 \left( \frac{3{\bf w}}{r^2} + \frac{2{\bf
m}}{r^3} + \frac{6{\bf k}}{r^4} \right) \eta_{AB} \]
\[ H_{33}=\frac 12 W_{03AB}\varepsilon^{AB}=
 \frac{3{\bf q}}{r^2} + \frac{2{\bf b}}{r^3} +
\frac{6{\bf s}}{r^4}  \]
\[ H_{3A}=W_{3B03}\varepsilon^B{_A}= \frac{3}{2r}{\bf q},_A + \frac{3}{r^2}
\varepsilon_{A}{^C}{\bf p},_C - \frac{3}{r^3}{\bf s},_A  \]
\[ H_{AB}=W_{0A3C}\varepsilon^C{_B}
= - \frac 12 \left( \frac{3{\bf q}}{r^2} + \frac{2{\bf b}}{r^3}
+ \frac{6{\bf s}}{r^4} \right) \eta_{AB} \]

\section{Conformal Yano--Killing tensors}

Let $Q_{\mu\nu}$ be an antisymmetric tensor field. Contracting the  Weyl tensor
$W^{\mu\nu\kappa\lambda}$ with $Q_{\mu\nu}$ we obtain a natural object
which can be integrated over two--surfaces. The result does not depend on the
choice of the surface if $Q_{\mu\nu}$ fulfills the following condition
introduced by Penrose
(see \cite{Pen-Rin} and \cite{JNG}):
\begin{equation}\label{Q}
Q_{\lambda (\kappa ;\sigma)} -Q_{\kappa (\lambda ;\sigma)} +
\eta_{\sigma[\lambda} Q_{\kappa ]}{^\delta}_{;\delta} =0
\end{equation}

We can rewrite equation (\ref{Q}) in a generalized form for
$n$--dimensional spacetime with metric $g_{\mu\nu}$:
\begin{equation}\label{Qn}
Q_{\lambda (\kappa ;\sigma)} -Q_{\kappa (\lambda ;\sigma)} +
\frac 3{n-1} g_{\sigma[\lambda} Q_{\kappa ]}{^\delta}_{;\delta} =0
\end{equation}
It is easy to check that equation (\ref{Qn}) is
equivalent to the following one:
\begin{equation}\label{CYK}
Q_{\lambda \kappa ;\sigma} +Q_{\sigma \kappa ;\lambda} =
\frac{2}{n-1} \left( g_{\sigma \lambda}Q^{\nu}{_{\kappa ;\nu}} +
g_{\kappa (\lambda } Q_{\sigma)}{^{\mu}}{_{ ;\mu}} \right)
\end{equation}
The tensor which fulfills the last equation will be called a
{\em conformal Yano
Killing tensor} (or simply CYK). The CYK tensor is a natural ``conformal''
generalization of the Yano tensor (see \cite{BF} and \cite{GR}).
More precisely,
for any scalar function $f>0$ and for a given metric $g_{\mu\nu}$
equation (\ref{CYK}) for $Q_{\mu\nu}$ and $g_{\mu\nu}$ is equivalent to
the same equation for $f^3 Q_{\mu\nu}$
and the conformally related metric $f^2 g_{\mu\nu}$.
 It is interesting to notice, that the
``square'' $A_{\mu\nu}$ of our tensor $Q_{\mu\nu}$:
\[ A_{\mu\nu} := Q_{\mu}{^\lambda}Q_{\lambda \nu} \] \noindent
fulfills the following equation:
\begin{equation}\label{Kt}
A_{(\mu\nu;\kappa)} = \frac 4{n-1} g_{(\mu\nu} Q_{\kappa)}{^\lambda}
Q_\lambda{^\delta}{_{ ; \delta}}
\end{equation}
which simply means that the symmetric tensor $A_{\mu\nu}$ is a conformal
Killing tensor.

For our purposes
we need to specify the formulae (\ref{Qn}) and (\ref{CYK})
to the
special case of the flat four--dimensional Minkowski space
($g_{\mu\nu}=\eta_{\mu\nu}$, $n=4$). In this simple situation the general
solution of (\ref{CYK}) or (\ref{Q})  assumes the following form in cartesian
coordinates $(x^\mu )$:
 \begin{equation}\label{Qs}
Q^{\mu\nu} = q^{\mu\nu} + 2u^{[ \mu}x^{\nu ]} -\varepsilon^{\mu\nu}
{_{\kappa\lambda}} v^{\kappa}x^{\lambda} -\frac 12 k^{\mu\nu} x_{\lambda}
x^{\lambda} +2k^{\lambda [\nu} x^{\mu ]} x_{\lambda}
\end{equation}
where $q^{\mu\nu}$, $k^{\mu\nu}$ are constant antisymmetric tensors and
$u^{\mu}$, $v^\mu$ are constant vectors.

It is easy to verify that the charge given by $Q_{\mu\nu}$ is well
defined. Indeed, we have:
\[ \int_{\partial V} W^{\mu\nu\lambda\kappa}Q_{\lambda\kappa}
 {\rm d}\sigma_{\mu\nu} =
\int_{V}
( W^{\mu\nu\lambda\kappa}Q_{\lambda\kappa}),_{\nu}
{\rm d}\Sigma_{\mu} = \]
\[ =
\int_{V} ( W^{\mu\nu\lambda\kappa},_{\nu} Q_{\lambda\kappa})+
W^{\mu\nu\lambda\kappa}Q_{\lambda\kappa},_{\nu} )
{\rm d}\Sigma_{\mu} = 0 \] \noindent
where the first term vanishes because of the field equations and
the second term vanishes because of the symmetries of the Weyl
tensor and because of equation
(\ref{Q}). The above equality implies that the flux of the quantity
$(W^{\mu\nu\lambda\kappa}Q_{\lambda\kappa})$ through any two closed
two--surfaces $S_1$ and $S_2$ is the same if there is a three--volume $V$
between them (i.e. if $\partial V = S_1 - S_2$). We define the charge
corresponding to the specific CYK tensor $Q$ as the value of this flux.

The above construction can be applied also to the dual ${^*} W$.
It turns out that we do not get more charges from ${^*} W$
because the dual $Q^*$ has the same form (\ref{Qs}) with the following
interchange:
\[ q \longleftrightarrow q^* \quad k \longleftrightarrow k^*
\quad u \longleftrightarrow v  \] \noindent
where $(Q^*)_{\mu\nu} := \frac 12 \varepsilon_{\mu\nu}{^{\lambda\kappa}}
Q_{\lambda \kappa}$.

Let us observe that the solutions (\ref{Qs}) form a
twenty--dimensional vector space. This means
that we have obtained 20 charges but only 14 remain when we pass to the
limit at spatial infinity assuming that $W \sim \frac 1{r^3}$ (the charge
related to $q$ is identically zero).
{\em I don't know any local argument
(i.e. using only field equations) for the vanishing of this charge.}

Let $\cal D$ be a generator of dilatations in Minkowski space.
We have the following commutation relations between generators of
pseudo--similarity group (Poincar\'e group extended by scaling transformation):
\[ {\cal T}_{\mu}=\base x\mu \; , \; \; \; {\cal L}_{\mu\nu}=x_{\mu} \base
x\nu - x_\nu \base x\mu \; , \; \; \; {\cal D}=x^\nu \base x\nu \]
\[ [{\cal T}_\mu , {\cal T}_\nu]=0 \]
\[ [{\cal T}_\mu, {\cal L}_{\alpha\beta}]=\eta_{\mu\alpha}{\cal T}_\beta
-\eta_{\mu\beta}{\cal T}_\alpha \]
\[ [{\cal T}_\mu , {\cal D}]={\cal T}_\mu \]
\[ [{\cal D}, {\cal L}_{\alpha\beta}] = 0 \]
\[
[{\cal L}_{\mu\nu}, {\cal L}_{\alpha\beta}] =
\eta_{\mu\alpha}{\cal L}_{\beta\nu} -\eta_{\mu\beta}{\cal L}_{\alpha\nu}+
\eta_{\nu\alpha}{\cal L}_{\mu\beta}-\eta_{\nu\beta}{\cal L}_{\mu\alpha} \]

As we already know the charges ${\bf k}, {\bf s}$ contain the information
about $j^{\mu\nu}$, ${\bf m},{\bf p}$ form a four--vector $p^\mu$ and
similarly ${\bf b},{\bf d}$ form a dual four--vector $b^\mu$.
 More precisely the following relations hold:

\centerline{ {$\displaystyle
 16\pi w_{\mu\nu}:=\int_{\partial\Sigma} W({\cal T}_\mu\wedge{\cal T}_\nu) $}}
\[
16\pi w_{z0}= 2\int_{S(r)}\Lambda W^{03}{_{z0}} =
-\int_{S(r)} \frac{\cos\theta}{r}(\Lambda \overline{\bf x}),_3 \]
\[
16\pi w_{xy}= 2\int_{S(r)}\Lambda W^{03}{_{xy}} =
-\int_{S(r)} \frac{\cos\theta}{r}(\Lambda \overline{\bf y}),_3 \]

\centerline{ {$\displaystyle  16\pi{^*}w_{\mu\nu} :=
 \int_{\partial\Sigma} {^*}W({\cal T}_\mu\wedge{\cal T}_\nu) =
 \int_{\partial\Sigma} W^*({\cal T}_\mu\wedge{\cal T}_\nu) $}}

\[ 16\pi {^*}w_{z0}= 2\int_{S(r)}\Lambda {^*}W^{03}{_{z0}} =
\int_{S(r)}\Lambda\cos\theta \varepsilon^{AB} W_{AB30} + \]
\[
+\int_{S(r)}\Lambda r\cos\theta \varepsilon^{AB} W_{AB0C||D}\eta^{CD}
=-\int_{S(r)} \frac{\cos\theta}{r}(\Lambda \overline{\bf y}),_3 \]
\centerline{ {$\displaystyle
16\pi p_\mu :=\int_{\partial\Sigma} W({\cal D}\wedge{\cal T}_\mu) $}}
\[
16\pi p_{0}= 2\int_{S(r)}\Lambda x^\mu W^{03}{_{\mu 0}} =
-\int_{S(r)} \frac{\Lambda}{r} \overline{\bf x} \]
\[ 16\pi p_{z}= 2\int_{S(r)}\Lambda x^\mu W^{03}{_{\mu z}} =
2\int_{S(r)}\Lambda t W^{03}{_{0z}} + \Lambda rW^{03}{_{3z}}=
16\pi t w_{0z} + \int_{S(r)}\overline{\bf X}\cos\theta  \]
\centerline{ {$\displaystyle
16\pi b_\mu :=\int_{\partial\Sigma} {^*}W({\cal D}\wedge{\cal T}_\mu) $}}
 \[
16\pi b_{0}= 2\int_{S(r)}\Lambda x^\mu {^*}W^{03}{_{\mu 0}} =
- \int_{S(r)}\Lambda r \varepsilon^{AB} W_{AB30} =
\int_{S(r)} \frac{\Lambda}{r} \overline{\bf y} \]
\[ 16\pi b_{z}= 2\int_{S(r)}\Lambda x^\mu {^*}W^{03}{_{\mu z}} =
2\int_{S(r)}\Lambda t {^*}W^{03}{_{0z}} + \Lambda r{^*}W^{03}{_{3z}}=
16\pi t {^*}w_{0z} + \int_{S(r)}\overline{\bf Y}\cos\theta  \]

\centerline{ {$\displaystyle
16\pi j_{\mu\nu} :=\int_{\partial\Sigma}
W({\cal D}\wedge{\cal L}_{\mu\nu}
-\frac 12 \eta({\cal D},{\cal D}) {\cal T}_\mu\wedge{\cal T}_\nu ) $}}

\[ 16\pi j_{0z}= 2\int_{S(r)}\Lambda x^\mu \left(rW^{03}{_{0\mu}}\cos\theta
+tW^{03}{_{z\mu}} \right) -\int_{S(r)}\Lambda x^\mu x_\mu W^{03}{_{0z}}= \]
\[ = -8\pi(r^2-t^2) w_{0z}  -16\pi t p_z +
 \int_{S(r)} \Lambda  \overline{\bf x}\cos\theta  \]

\[ 16\pi j_{xy}= 2\int_{S(r)}\Lambda x^\mu \left(
 W^{03}{_{\mu y}} x -  W^{03}{_{\mu y}} y \right) -
 \int_{S(r)}\Lambda x^\mu x_\mu  W^{03}{_{0z}}= \]
\[ = -8\pi(r^2-t^2) w_{xy}  -16\pi t b_z +
 \int_{S(r)}  \Lambda  \overline{\bf y}\cos\theta \]

\centerline{ {$\displaystyle
16\pi {^*}j_{\mu\nu} :=\int_{\partial\Sigma}
{^*}W({\cal D}\wedge{\cal L}_{\mu\nu}
-\frac 12 \eta({\cal D},{\cal D}) {\cal T}_\mu\wedge{\cal T}_\nu ) =
\int_{\partial\Sigma}
W^*({\cal D}\wedge{\cal L}_{\mu\nu}
-\frac 12 \eta({\cal D},{\cal D}) {\cal T}_\mu\wedge{\cal T}_\nu ) $}}
\vspace{1ex}

\noindent The conservation law for the charge $w_{\mu\nu}$ is a consequence
of field equations:
\[ \int_{\partial\Sigma} W^{\mu\nu}{_{\lambda\kappa}}
 {\rm d}\sigma_{\mu\nu} =
\int_{\Sigma} (W^{\mu\nu}{_{\lambda\kappa}}),_{\nu}
{\rm d}\Sigma_{\mu} =0 \] \noindent
For $p^\mu$ and $b^\mu$ we obtain the conservation laws from the
following observation:
\[ \int_{\partial\Sigma} x^{\lambda}W^{\mu\nu}{_{\lambda\kappa}}
 {\rm d}\sigma_{\mu\nu} =
\int_{\Sigma} \left( x^{\lambda}W^{\mu\nu}{_{\lambda\kappa}},_{\nu} +
\delta^{\lambda}_\nu W^{\mu\nu}{_{\lambda\kappa}} \right)
{\rm d}\Sigma_{\mu} =0 \] \noindent
(the same holds for ${^*}W$).\\
For $j^{\mu\nu}$ the corresponding identities are as follows:
\[ \int_{\partial\Sigma} (x_{\lambda}W^{\mu\nu\lambda\kappa}x^{\delta}
-x_{\lambda}W^{\mu\nu\lambda\delta}x^{\kappa}
 -\frac 12 x^{\lambda}x_{\lambda} W^{\mu\nu\delta\kappa})
 {\rm d}\sigma_{\mu\nu} = \]
 \[ =\int_{\Sigma} (x_{\lambda}W^{\mu\nu\lambda\kappa}x^{\delta}
-x_{\lambda}W^{\mu\nu\lambda\delta}x^{\kappa}
 -\frac 12 x^{\lambda}x_{\lambda} W^{\mu\nu\delta\kappa} ),_{\nu}
{\rm d}\Sigma_{\mu} = \]
\[ =
\int_{\Sigma} (x_{\lambda}W^{\mu\delta\lambda\kappa}
-x_{\lambda}W^{\mu\kappa\lambda\delta} - x_{\lambda}
W^{\mu\lambda\delta\kappa} ){\rm d}\Sigma_{\mu} =  \]
\[ =\int_{\Sigma} x_{\lambda}
(W^{\mu\delta\lambda\kappa}+W^{\mu\kappa\delta\lambda} +
W^{\mu\lambda\kappa\delta}){\rm d}\Sigma_{\mu} =0\]

\section{Asymptotically flat spacetimes}

Consider an asymptotically flat spacetime (at spatial infinity),
fulfilling the (complete non-linear) Einstein equations. Suppose,
moreover, that the energy--momentum tensor of the matter vanishes
around spatial infinity (``sources of compact support'').
This means that
the Riemann tensor and the Weyl tensor do coincide outside of the world
tube containing the matter. Let us analyze, for simplicity, this
situation in terms of an asymptotically flat coordinate system
(for nice geometric formulations of
asymptotic flatness see e.g. \cite{AH} or \cite{AR}).
We suppose that there exists an (asymptotically
Minkowskian) coordinate system $(x^\mu)$:
\[ g_{\mu\nu} - \eta_{\mu\nu} \sim r^{-b} \; \; \; \; \;
 g_{\mu\nu,\lambda} \sim r^{-b -1} \] \noindent
where $\displaystyle r:= \sum_{k=1}^{3} (x^k)^2$ and typically $b=1$ (but
$1\geq b > \frac 12$ is also possible -- see \cite{Ch1}).

For a general asymptotically flat (AF) metric we cannot expect that the
equations (\ref{Q}) and
(\ref{CYK}) admit any solution. Instead, we assume that the
left--hand side of (\ref{Q}): \[ {\cal Q}_{\lambda\kappa\sigma}:=  Q_{\lambda
(\kappa ;\sigma)} -Q_{\kappa (\lambda ;\sigma)} +
 g_{\sigma[\lambda} Q_{\kappa ]}{^\delta}_{;\delta}  \] \noindent
has a certain asymptotic behaviour at spatial infinity
\begin{equation}\label{QAF}
 {\cal Q}_{\mu\nu\lambda} \sim r^{-c}  \end{equation}
 On the other hand, suppose that  $Q_{\mu\nu}$ behaves asymptotically as
follows:
 \[ Q_{\mu\nu} \sim r^{a} \; \; \; \; \;
 Q_{\lambda \kappa , \sigma} \sim r^{a-1}  \]
\noindent
Moreover, suppose that the Riemann tensor $R_{\mu\nu\kappa\lambda}$
behaves asymptotically as follows:
\[ R_{\mu\nu\kappa\lambda} \sim r^{-b -1 -d} \] \noindent
It can be easily checked (see e.g. \cite{JNG}) that the vacuum Einstein
equations imply the following equality:
 \begin{equation}\label{*R*}
 \nabla_{\lambda} \left( {^*\!}R{^*}^{\mu\lambda}{_{\alpha\beta}}
Q^{\alpha\beta} \right) = - \frac 23
R^{\mu\lambda \alpha\beta} Q_{\lambda\alpha\beta} \end{equation}

The left--hand side of (\ref{*R*}) defines an asymptotic charge provided
that the right--hand side vanishes sufficiently fast at infinity. It is easy to
check that, for this purpose, the exponents $b,c,d$ have to fulfill the
inequality
 \begin{equation}\label{2} b + c + d > 2  \end{equation}
In typical
situation when $b=d=1$, the inequality (\ref{2}) simply means that $c > 0$. In
this case a weaker condition is also possible (for example ${\cal
Q}_{\mu\nu\lambda} \sim (\ln r)^{-1-\epsilon}$ with $\epsilon >0$).

Let us define an {\em asymptotic conformal Yano--Killing tensor} (ACYK) as an
antisymmetric
tensor $Q_{\mu\nu}$ such that ${\cal Q}_{\mu\nu\lambda} \rightarrow
0$ at spatial infinity.
For constructing the ACYK tensor we can begin with the solutions (\ref{Qs})
in flat Minkowski space. Asymptotic behaviour at infinity of these flat
solutions explain why we expect for any ACYK tensor the following
behaviour:
\[ Q_{\mu\nu} = {^{(2)}\!}Q_{\mu\nu} + {^{(1)}\!}Q_{\mu\nu} +
{^{(0)}\!}Q_{\mu\nu} \] \noindent
where ${^{(2)}\!}Q_{\mu\nu} \sim r^2$, ${^{(1)}\!}Q_{\mu\nu} \sim r$ and
${^{(0)}\!}Q_{\mu\nu} \sim r^{1-c}$.

It is easy to verify that $c \geq b +1 -a$ and if $b=1$
than for $a=2$ we have $c \geq 0$. This means that in a general situation
there are no solutions of (\ref{QAF}) with nontrivial
${^{(2)}\!}Q_{\mu\nu}$ and $c>0$.
This is the origin of the difficulties with the definition of the angular
momentum.
On the other hand it is easy to check that the energy--momentum four--vector
and the dual one are well defined ($a=c=1$)  and the condition
$b + d > 1$ can be easily fulfilled (typically $b=d=1$).

\section{Strong asymptotic flatness}

Here, we propose a new, stronger definition of the asymptotic flatness.
The definition is motivated by the above discussion.

Suppose that there exists a coordinate system $(x^\mu)$
such that:
\[ g_{\mu\nu} - \eta_{\mu\nu} \sim r^{-1} \]
\[ \Gamma^{\kappa}{_{\mu\nu}} \sim r^{-2} \]
\[ R_{\mu\nu\kappa\lambda} \sim r^{-3} \] \noindent
In the space of ACYK tensors fulfilling the asymptotic condition
\begin{equation}\label{Q1}
 Q_{\lambda (\kappa ;\sigma)}
-Q_{\kappa (\lambda ;\sigma)} +
g_{\sigma[\lambda} Q_{\kappa ]}{^\delta}_{;\delta} =
{\cal Q}_{\lambda\kappa\sigma} \sim r^{-1}  \end{equation}
we define the following equivalence relation:
\begin{equation}\label{rel} Q_{\mu\nu} \equiv Q_{\mu\nu}'
\Longleftrightarrow Q_{\mu\nu}  - Q_{\mu\nu}' =  O(1)
 \end{equation}
for $r \rightarrow \infty$.
We assume that the space of equivalence classes defined by (\ref{Q1}) and
(\ref{rel}) has a finite dimension $D$ as a vector space.
The maximal dimension $D = 14$ correspond to the situation where there
are no supertranslation problems
in the definition of an angular momentum.
In the case of spacetimes for which $D<14$
the lack of certain ACYK tensor means that the corresponding charge
is not well defined.

\section*{Acknowledgments}
The author is much indebted to Ch. Duval, J.
Kijowski and R. Kerner for fruitful discussions and to
the Laboratoire de Gravitation et Cosmologie Relativistes, Universit\'e
Paris VI and the Centre de Physique Th\'eorique in Marseille (Luminy)
for the hospitality during the preparation of this paper. Also, many
thanks to the referee for his remarks and observations which were used to
improve the final presentation of the results.

The author is very much indebted for the financial support, which he
got from the European Community (HCM Contract No. CIPA--3510--CT92--3006)
and from the Polish National Committee for Scientific Research (Grant No.
2 P302 189 07).

\section*{Appendix --- linear ADM equations}

Linear Einstein equations are as follows:
\[
2\Lambda^{-1}\dot{P}^{33}= -h^0{_0}{^{||A}}_A -2r^{-1}h^0{_0},{_3} +
 h^{33 ||A}{_A} + 2r^{-2}h^{33} + 16\pi T^{33} + \]
\begin{equation} +  (H,{_3}- 2h^{3A}{_{||A}} -2r^{-1}h^{33}),{_3}
+2r^{-1}(H,{_3}-2h^{3A}{_{||A}}-2r^{-1}h^{33}) 	\label{P33}
\end{equation}
\[
2\Lambda^{-1}\dot{P}^{3C}= (h^0{_0},{_3} -r^{-1}h^0{_0})^{ ||C} +
\frac{1}{2} (H,{_3}- 2h^{3A}{_{||A}} -2r^{-1}h^{33})^{ ||C} -
 \chi^A{_{B||A}},{_3}\eta^{CB} +\]
\begin{equation}
+h^{3C||A}{_{||A}} +h^{3A||C}{_{||A}}-h^{3A}{_{||A}}^{||C}
+ 16\pi T^{3C} \label{P3C}
\end{equation}
\[
2\Lambda^{-1}\dot{P}_{AB} = h^0{_{0||AB}} + \eta_{AB}(r^{-1}h^0{_0},{_3}
 - h^0{_0}{^{|m}}_m) +
[\eta_{AB}(\frac{1}{2} H,{_3}- h^{3A}{_{||A}} -r^{-1}h^{33})],{_3} +
\]
\[ + 2r^{-1}\eta_{AB}(\frac{1}{2} H,{_3}- h^{3C}{_{||C}} -r^{-1}h^{33}) +
(\frac{1}{2}H_{||C}{^C}+r^{-2}H)\eta_{AB}+\]
\[ - (h^3{_{A||B}}+h^3{_{B||A}}-
\eta_{AB}h^{3C}{_{||C}}),{_3}+ h^{33}{_{||AB}} + r^{-2}\eta_{AB}h^{33} +
\]
\begin{equation}
+ (\chi^C{_B},{_3}\eta_{CA}),{_3}
 + \chi_{AB}{^{||C}}_{||C} -
\chi^C{_{A||BC}} - \chi^C{_{B||AC}} + 16\pi T_{AB}  \label{PAB}
\end{equation}
\begin{equation}
\dot{h}_{33} = \Lambda^{-1}(P^{33}-S) + 2h_{03},{_3}	\label{h33}
\end{equation}
\begin{equation}
\dot{h}^{3A} =2\Lambda^{-1}P^{3A} + h_{03}{^{||A}} + h_0{^A},{_3}   \label{h3A}
\end{equation}
\begin{equation}
\dot{h}_{AB}
= 2\Lambda^{-1}S_{AB}- \eta_{AB}\Lambda^{-1}P_{33} +h_{0A||B} + h_{0B||A} +
 2r^{-1} \eta_{AB} h_{03} 	\label{hAB}
\end{equation}
\[
 h ^{|l}{_l} - h^{kl}{_{|kl}} = (H,{_3}-2h^{3A}{_{||A}}-2r^{-1}h^{33}),{_3} +
3r^{-1}(H,{_3}-2h^{3A}{_{||A}}-2r^{-1}h^{33})+\]
\begin{equation}
 +h^{33 ||A}{_A}+2r^{-2}h^{33}+(\frac{1}{2}H^{||C}{_C}+r^{-2}H)
  - \chi^{AB}{_{||AB}} = -16\pi T^{00}   \label{wsk}
\end{equation}
\begin{equation}
P^{33},{_3} +  P^{3A}{_{||A}} - r^{-1}S = 8\pi \Lambda T_{03}  \label{ww3}
\end{equation}
\begin{equation}
P_{3A},{_3} + P_A{^B}{_{||B}} =
 P_{3A},{_3} + S_A{^B}{_{||B}}+\frac{1}{2} S_{||A} = 8\pi\Lambda T_{0A}
\label{wwA}
\end{equation}


\begin{thebibliography}{666}

\bibitem{Dirac} P.A.M. Dirac, Proc. Roy. Soc. A133, p. 60 (1931); Phys.
Rev. 74, p. 817 (1948)

\bibitem{Amaldi} E. Amaldi, {\it On the Dirac Magnetic Poles}, in: Old and
New Problems in Elementary Particles, ed. G. Puppi, Academic Press, (New
York 1968)

\bibitem{ADM} R. Arnowitt, S. Deser, C. Misner, {\it The dynamics of general
relativity}, in: Gravitation: an introduction to current research,
ed. L. Witten, p. 227, Wiley, (New York 1962)

\bibitem{AT} A. Trautman, {\it Conservation laws in general
relativity}, in: Gravitation: an introduction to current research,
ed. L. Witten, p. 169, Wiley, (New York 1962)

\bibitem{MTW} C. Misner, K.S. Thorne, J. A. Wheeler {\it Gravitation},
W.H. Freeman and Co., (San Francisco 1973)

\bibitem{GRG} J. Jezierski and J. Kijowski, General Relativity and
Gravitation 22,  1283--1307 (1990)

\bibitem{Ch-Kl} D. Christodoulou, S. Klainerman, Communications on Pure
and Applied Mathematics 43,  137--199 (1990)

\bibitem{NP} E.T. Newman, R. Penrose, Phys. Rev. Lett. 15, 231 (1965);
Proc. Roy. Soc. A305, 175--204 (1968)


\bibitem{Pen-Rin} R. Penrose and W. Rindler, {\it Spinors and Space-time},
Cambridge University Press, Vol. 2, p.396 (Cambridge 1986)

\bibitem{JNG} J.N. Goldberg, Physical Review D 41, 410--416 (1990)

\bibitem{AR} A. Ashtekar, J.D. Romano, Classical and Quantum Gravity,
1069--1100 (1992)

\bibitem{CZ} A. Cresswell, R.L. Zimmerman, General Relativity and
Gravitation 20, 927--941 (1988)

\bibitem{NUT} E.T. Newman, L. Tamburino, T.J. Unti, Journal of Mathematical
Physics 4, 915--923 (1963)

\bibitem{MD} M. Demia\'nski, Physics Letters A42, p.157 (1972)

\bibitem{AK} A. Krasi\'nski, {\it Physics in an Inhomogeneous Universe (a
review)}, Warsaw University Publishers, (Warsaw 1993)

\bibitem{BF} S. Benenti, M. Francaviglia, General Relativity and Gravitation,
Edited by A. Held, vol. 1, 393--439, Plenum Press, (New York 1980)

\bibitem{GR} G.W. Gibbons, P.J. Ruback, Commun. Math. Phys. 115, 267--300
(1988)

\bibitem{AH} A. Ashtekar, R.O. Hansen, Journal of Mathematical
Physics 19, 1542--1566 (1978)

\bibitem{TF} T. Friedrich, {\it Self--duality of Riemannian Manifolds
and Connections}, in: Self--dual Riemannian Geometry and Instantons,
TEUBNER--TEXTE zur Mathematik, Band 34, (Leipzig 1981)

\bibitem{BB} I. Bia\,lynicki--Birula and Z. Bia\,lynicka--Birula,
{\it Quantum Electrodynamics}, Pergamon, (Oxford 1975)

\bibitem{Ch1} P.T. Chru\'sciel, Communications in
Mathematical Physics 120, 233--248 (1988)

\end{thebibliography}
\end{document}